\documentclass[aps,twocolumn,altaffilletter,nolongbibliography,numerical,flushbottom,secnumarabic,pra,superscriptaddress,floatfix,10pt]{revtex4-1}

\usepackage{graphicx}
\usepackage{upgreek}
\usepackage{amsmath, amssymb}
\usepackage{braket}
\usepackage{newtxtext, newtxmath}
\usepackage{siunitx}
\usepackage{booktabs}

\usepackage[breaklinks, pdftex, hyperfootnotes=true, pdfpagelabels, bookmarks, pageanchor]{hyperref}
\hypersetup{%
	colorlinks=true, linktocpage=true, pdfstartpage=1, pdfstartview=FitH, pdfborder={0 0 0},%
	breaklinks=true, pdfpagemode=UseNone, pageanchor=true, pdfpagemode=UseOutlines,%
	plainpages=false, bookmarksnumbered, bookmarksopen=true, bookmarksopenlevel=1,%
	hypertexnames=true, pdfhighlight=/O,
	urlcolor=blue, linkcolor=blue, citecolor=blue, 
}

\usepackage{lipsum}

\newcommand{\ie}{i.\,e.}

\newcommand{\hc}{\text{h.\,c.}}
\newcommand{\ped}[1]{_\text{#1}}
\newcommand{\api}[1]{^\text{#1}}
\newcommand{\comm}[2]{\left[#1, #2\right]}
\newcommand{\nspin}{n}
\newcommand{\n}{N}
\newcommand{\ham}{H}
\newcommand{\hamtf}{\ham\ped{q}}
\newcommand{\hampr}{\ham\ped{p}}
\newcommand{\hamcd}{\ham\ped{cd}}

\newcommand{\tf}{T}

\newcommand{\mingap}{\Delta}
\newcommand{\pgs}{P\ped{gs}}

\newcommand{\iu}{i}
\newcommand{\eu}{e}

\DeclareMathOperator{\Tr}{Tr}
\DeclareMathOperator{\dist}{dist}

\begin{document}

\title{Counterdiabatic driving in the quantum annealing of the $p$-spin model: a variational approach}	

\author{G.\,Passarelli}
\affiliation{Dipartimento di Fisica ``E.\,Pancini'', Universit\`a di Napoli Federico II, Complesso di Monte S.~Angelo, via Cinthia - 80126 - Napoli, Italy}
\affiliation{CNR-SPIN, c/o Complesso di Monte S. Angelo, via Cinthia - 80126 - Napoli, Italy}
\author{V.\,Cataudella}
\affiliation{Dipartimento di Fisica ``E.\,Pancini'', Universit\`a di Napoli Federico II, Complesso di Monte S.~Angelo, via Cinthia - 80126 - Napoli, Italy}
\affiliation{CNR-SPIN, c/o Complesso di Monte S. Angelo, via Cinthia - 80126 - Napoli, Italy}
\author{R.\,Fazio}
\affiliation{Abdus Salam  ICTP, Strada Costiera 11, I-34151 Trieste, Italy}
\affiliation{Dipartimento di Fisica ``E.\,Pancini'', Universit\`a di Napoli Federico II, Complesso di Monte S.~Angelo, via Cinthia - 80126 - Napoli, Italy}
\author{P.\,Lucignano}
\affiliation{Dipartimento di Fisica ``E.\,Pancini'', Universit\`a di Napoli Federico II, Complesso di Monte S.~Angelo, via Cinthia - 80126 - Napoli, Italy}
\affiliation{CNR-SPIN, c/o Complesso di Monte S. Angelo, via Cinthia - 80126 - Napoli, Italy}

\begin{abstract}
 Finding the exact counterdiabatic potential is, in principle, particularly demanding. 
 Following recent progresses about variational strategies to approximate the counterdiabatic operator, in this 
paper we apply this technique to  the quantum annealing of the $p$-spin model. In particular, for $ p = 3 $ we find a new form of the counterdiabatic  potential originating from a cyclic ansatz, that allows us to have optimal fidelity even for extremely short dynamics, independently of the size of the system.  We compare our 
results with a nested commutator ansatz, recently proposed in P.\,W.~Claeys, M.\,Pandey, D.\,Sels, and A.\,Polkovnikov, Phys.\,Rev.\,Lett.\,123,~090602~(2019), for $p=1$ and $p=3$. We also analyze generalized $ p $-spin models to get a further insight into our ansatz.
\end{abstract}

\maketitle	

\section{Introduction}
It is possible to solve a complex optimization problem by finding the ground state of a Hamiltonian $ \hampr $ that encodes the problem of interest.
Adiabatic quantum computation~\cite{albash:aqc} and quantum annealing~\cite{santoro-review, hauke-rev} achieve this task by slowly dragging the 
system to the desired state, having first initialized it in the ground state of a known (simple) Hamiltonian $\hamtf$. The adiabatic theorem assures 
that a time-dependent Hamiltonian $\ham_0(t)$,  such that $ \ham_0(0) = \hamtf $ and  $ \ham_0( \tf ) = \hampr $, will lead towards the wanted  solution if, at $t=0$, the system is prepared in the ground state of $ \ham_0(0)$ and if $ \tf $ is large enough to avoid transitions to excited 
states.

The ultimate bound to the efficiency of this approach is  related to the presence, and to the scaling with system size,  of vanishing small gaps in the 
instantaneous spectrum of $\ham_0(t)$. The evolution time must be tuned so that at $ t = \tf $ there is a large probability of finding the system in its 
ground state (longer annealing times obviously leave also the system more prone to errors and thermal noise~\cite{albash:decoherence}). According 
to the adiabatic criterion, this means that $ \tf $ must be longer than $ \hslash / \mingap^2 $, where $ \mingap $ is the minimal gap between the ground 
state and the first excited state. When the spectral gap closes in the thermodynamic limit, as in the case of a quantum phase transition~\cite{sachdev:qpt}, 
the evolution time $ \tf $ must grow with the system size as well, to compensate for the closure of the minimal gap. The hardness of a given optimization 
problem is directly reflected  in the presence of exponentially small (with the system size) gaps during the adiabatic evolution.
		
Finding ways to reduce the computation time $ \tf $, while keeping the fidelity of the time-evolved state (at $t= \tf $) close to the ground state of $ \hampr $,
is a central question in this context that still needs an adequate answer. In the last years, there have been several attempts to circumvent this problem, resorting to quantum optimal control approaches applied to the evolution of a complex system. A particularly appealing approach that 
received increasing attention recently is the implementation of shortcuts to adiabaticity~\cite{Demirplak:2003,Berry2009,torrentegui:sta} to quantum 
annealing.

A Shortcut To Adiabaticity (STA) [or Counterdiabatic Driving (CD)] amounts in finding a time-dependent Hamiltonian which evolve the state of a system 
as if the evolution was adiabatic. Given the bare Hamiltonian $ \ham_0(t) $, it is possible to find a term $\hamcd(t)$ such that that the dynamics governed   
by $ \ham(t) = \ham_0(t)+\hamcd(t)$  has vanishing diabatic transitions between pairs of energy eigenstates at all times~\cite{Berry2009}. Since its initial formulation, 
STAs have found numerous applications. An overview of the field can be found in Ref.~\cite{STA-rev}. 

The exact CD Hamiltonian $\hamcd(t)$, derived by Berry~\cite{Berry2009}, can be expressed in terms of the exact instantaneous eigenstates and eigenvalues 
of $ \ham_0(t)$. In a many-body system, $\hamcd(t)$ becomes progressively nonlocal and hence not easy for a practical implementation, when  $ \ham_0(t)$
is close to a quantum phase transition~\cite{delcampo-2012,fazio:sta}. More importantly from a conceptual point of view, the severe limitation in the use of the 
exact form of $\hamcd(t)$ arises because i) exact eigenstates cannot be determined (this would correspond to a prior knowledge of the solution of the optimization),
and ii) $\hamcd(t)$ may be ill-defined due to small denominators near the quantum critical point.  

A big step forward in overcoming these limitations has been achieved by \textcite{SelsPolkovnikov}, who derived a variational principle to determine approximate forms of $ \hamcd $. This new variational approach to STA solicits detailed scrutiny of its potentialities in speeding up quantum annealing protocols.  Some very recent papers addressed this question~\cite{Hartmann_2019, FloquetManyBody, hatomura:sta-mf}. \textcite{Hartmann_2019} analyzed the approximate optimal counterdiabatic driving in the so called LHZ lattice gauge model~\cite{lechner_2015}, \textcite{FloquetManyBody} studied a non-integrable one-dimensional Ising model, \textcite{hatomura:sta-mf} adopted a mean field approximation for the CD driving of the infinite-range Ising model.

In this work, we continue along this program and apply the variational approach by Sels and Polkovnikov to the quantum annealing of the ferromagnetic $ p $-spin model, which is currently subject of intense investigation~\cite{derrida:p-spin, gross:p-spin, bapst:p-spin, nishimori:p-spin-1,nishimori:p-spin-2,nishimori:p-spin-3,
nishimori:p-spin-4,nishimori:p-spin-5,nishimori:p-spin-6,nishimori:p-spin-7,passarelli:dissipative-p-spin,passarelli:pausing,acampora:genetic-embedding,
passarelli:reverse}.  Despite being exactly solvable, the $p$-spin model has a non-trivial phase diagram and is deeply related to 
NP-hard optimization.  Furthermore, it encodes a Grover-like search~\cite{grover:search,roland-cerf}  for large and odd $ p $. 

The paper is organized as follows. In Section~\ref{variational}, we  introduce the basic concepts  
of STA and describe the variational procedure developed in  Ref.~\cite{SelsPolkovnikov}.  In the same section, we show the approximate counterdiabatic operator 
adopting two different ans\"atze for the variational potential: The first one is based on the Nested Commutator (NC) approach~\cite{FloquetManyBody}, the second one is based on a simple Cyclic Ansatz (CA), the origin of this name will be clear in the following. In Section~\ref{pspin}, we discuss different properties of the $p$-spin model for $ p = \text{\numlist{1;2;3}} $ in detail. This section contains a detailed discussion of the results we obtained.  We compare the two approaches and show that, in the case $ p = 3 $, the CA outperforms the NC ansatz considerably, leading to an almost perfect recovery of the ground state in the final time $ \tf $. 
This effect occurs even for very large system sizes, the scaling of the fidelity being almost independent of the number of spins. 
Given the excellent performance of the CA for the $p$-spin model, in Section~\ref{non-pspin}, we move away from this highly-symmetric case and consider different generalizations that do not have constant all-to-all couplings. The question is to see if the CA has sufficient power and flexibility to be used in a generic situation. To this aim, we  consider different finite-range and random variants of the $ p $-spin model and compare the two variational ans\"atze also in these cases. 
Finally, in the Section~\ref{conclusions}, we summarize our results and briefly discuss future directions.

\section{Quantum annealing and counterdiabatic driving}
\label{variational}

As already mentioned in the introduction, a  typical annealing schedule can be formulated as follows. A (many-body) quantum system is 
governed by a time-dependent Hamiltonian of the form
		\begin{equation}
			\ham_0(\lambda(t)) = \left[1-\lambda(t)\right] \hamtf + \lambda(t) \hampr 
		\end{equation}
with the constraint  that $ \ham_0(0) = \ham(0) $ and $ \ham_0(\tf) = \ham(\tf) $, which  implies $\lambda(0) = 0,   \lambda(\tf) = 1 $. 
In the rest of the paper we will further impose that $ \dot{\lambda}(0) = \dot{\lambda}(\tf) = 0 $. A convenient parametrization that automatically 
includes all these requirements has the form
		\begin{equation}
			\lambda(t) = \sin^2 \left[\frac{\pi}{2} \sin^2\left( \frac{\pi t}{2 \tf}\right)\right].
		\end{equation}
The goal is to reach, as close as possible, the ground state of $ \ham(\tf) $  if the initial state is the ground state of  $ \ham(0) $.  Having 
introduced the instantaneous eigenstates of $\ham_0(\lambda(t))$ as $\ham_0(\lambda(t)) \ket{\epsilon_m(t)} = \epsilon_m \ket{\epsilon_m(t)}$,
ideally we require that 
		\begin{equation}
			 \ket{\psi(\tf)} = \ket{\epsilon_0(\tf)}.
		\label{psiT}
		\end{equation}
		
\subsection{Shortcut to adiabaticity}		
\label{STA-subsection}

Achieving the condition in Eq.~\eqref{psiT} is the goal of adiabatic quantum computation via quantum annealing. If one wants to speed 
up the evolution, an additional control on the dynamics is required. The proposals  collectively named shortcuts to adiabaticity or
counterdiabatic  drivings~\cite{Demirplak:2003,Berry2009,torrentegui:sta,delcampo:sta}, reviewed in Ref.~\cite{STA-rev},
are based on the idea that the evolution of a quantum state can coincide with the instantaneous ground state of $\ham_0(\lambda(t))$
if an additional time-dependent contribution $\hamcd(t)$ is added to the Hamiltonian. The exact quantum state  governed by the 
Hamiltonian $\ham_0(\lambda(t))+ \hamcd(t)$ is given by $\ket{\psi(t)} = \ket{\epsilon_0(t)}$ at any time (here, for simplicity, we considered 
the case in which one follows the ground state).

Berry~\cite{Berry2009} derived the exact CD potential to add to the bare Hamiltonian $ \ham_0(t) $ to completely suppress diabatic 
transitions between pairs of energy eigenstates at all times. It reads
		\begin{equation}\label{eq:cd-exact}
			\hamcd(t) \equiv \iu \hslash \sum_{m \ne l} \frac{\braket{\epsilon_m(t) | \partial_t \ham_0(t) | \epsilon_l(t)}}{\epsilon_l - \epsilon_m} 
			\ket{\epsilon_m(t)}\bra{\epsilon_l(t)}.
		\end{equation}
 Albeit exact, this potential is 
ill-defined at the quantum critical point due to exponentially small denominators.  Furthermore, even in the cases where Eq.~\eqref{eq:cd-exact} 
can be analytically computed, the resultant operator can be highly nonlocal and impossible to implement on the available hardware. 
Finally, and most important for its application to quantum annealing, the Hamiltonian in Eq.~\eqref{eq:cd-exact}  requires the knowledge of the exact 
eigenvalues and eigenvectors of the Hamiltonian at any time.  

A big step forward in offering a solution to these problems comes with the variational approach to counterdiabatic driving formulated 
in Refs.~\cite{SelsPolkovnikov,KOLODRUBETZ20171}.

\subsection{A variational approach to counterdiabatic driving}		
\label{variational-subsection}
		
A  variational principle for the counterdiabatic potential $ \hamcd $ can be derived as follows.
If one parametrizes  the CD potential as   $ \hamcd(t) = \dot{\lambda}(t) A_\lambda(t) $,  $ A_\lambda $ satisfies
		\begin{equation}
			\iu \hslash \left(\partial_\lambda \ham_0 + F\ped{ad}\right) = \comm{A_\lambda}{\ham_0},
		\end{equation}
with
		\begin{equation}
			F\ped{ad} = -\sum_m \partial_\lambda \epsilon_m \ket{\epsilon_m} \bra{\epsilon_m}.
		\end{equation}
It is possible to  introduce the operator~\cite{SelsPolkovnikov,KOLODRUBETZ20171}:
		\begin{equation}\label{eq:operator-G}
			G_\lambda = \partial_\lambda \ham_0 + \frac{\iu}{\hslash} \comm{A_\lambda^*}{\ham_0},
		\end{equation}
whose diagonal elements in the energy eigenbasis do not depend on $ A_\lambda^* $, and whose off-diagonal elements in the energy 
eigenbasis are zero if $ A_\lambda^* = A_\lambda $ (\ie, if $ G_\lambda = -F\ped{ad} $). Therefore, the true counterdiabatic potential 
$ A_\lambda $ minimizes the Hilbert-Schmidt norm of $ G_\lambda $, \ie, the operator distance between $ G_\lambda $ and $ -F\ped{ad} $:
		\begin{equation}\label{eq:action}
			\frac{\delta S(A_\lambda^*)}{\delta A_\lambda^*} \, \Bigg\rvert_{A_\lambda^* = A_\lambda} = 0, \qquad S(A_\lambda^*) 
			= \Tr[G_\lambda^2(A_\lambda^*)].
		\end{equation}

In the next Section we are going to apply the above mentioned variational approach to the $p$-spin model.
As in any variational approach, the next important step is an educated guess of (in this case) the operator $A_\lambda$. 
We will consider two variational  forms of the operator	 $A_\lambda$:
\begin{enumerate}
\item  An operator $A_\lambda^{(l)}$ as defined in Ref.~\cite{FloquetManyBody}, expressed in the form of NC:
		\begin{align}\label{eq:nested-commutators}
			A_\lambda^{(l)} &= \iu \hslash \sum_{k=1}^{l} \alpha_k [\underbrace{\ham_0, [\ham_0, \dots [\ham_0}_{2k-1}, \partial_\lambda \ham_0]]] 
			\notag \\
			&= \iu\hslash \sum_{k=1}^{l} \alpha_k O_{2k-1}.
		\end{align}
It turns out that this form can be manipulated efficiently. Important details of the method are discussed in Appendix~\ref{ansatz}.     In principle, the larger is $l$, the more accurate is the approximation of the CD potential.
\item  For $p=1,3$, we also study a new form, that we name CA, where the corresponding variational operator $A_\lambda\api{CA}$
	has a particularly symmetric form when applied to the model considered here. It is expressed by
\begin{equation}\label{eq:sy-ansatz}
	A_\lambda\api{CA} = \sum_{i=1}^{p'} \alpha_i S^i_y + \sum_{abc} \alpha_{abc} \epsilon_{a b c} S_a S_b S_c .
\end{equation} 
Details on this ansatz are discussed in Appendix~\ref{CA_appendix}. 
\end{enumerate}
In the next sections, we scrutinize these two possibilities testing to which extent it is possible to optimize the annealing schedule in these cases. 		
The new variational ansatz, especially tailored to this specific model, leads to exceptional good fidelities for the $p$-spin model.
		
\section{Ferromagnetic $p$-spin model}
\label{pspin}
		
The ferromagnetic $ p $-spin model is defined by the  Hamiltonian
		\begin{equation}\label{eq:p-spin}
			\hamtf = -2\Gamma S_x , \quad \hampr = -\frac{J}{\nspin^{p-1}} {(2S_z)}^p,
		\end{equation}
where $ S_x$, $S_y$, $S_z $ are total spin operators:
		\begin{equation}
			S_{j} = \frac{\hslash}{2}\sum_{i=1}^{\nspin} \sigma_i^{j}, \qquad j\in\set{x, y, z}.
		\end{equation}
For odd $ p $, the ground state of $ \hampr $ is ferromagnetic and 
nondegenerate, whereas for even $ p $ there are two degenerate ferromagnetic ground states, related by a $ Z_2 $ symmetry. 
In the following, we choose $ \Gamma = J = 1 $ and $ \hslash = 1 $.
		
We  consider three prototypical cases of this model~\cite{wauters:p-spin, derrida:p-spin, gross:p-spin, bapst:p-spin, nishimori:p-spin-1,
nishimori:p-spin-2,nishimori:p-spin-3,nishimori:p-spin-4,nishimori:p-spin-5,nishimori:p-spin-6,nishimori:p-spin-7,passarelli:dissipative-p-spin,passarelli:pausing,acampora:genetic-embedding,passarelli:reverse}:
\begin{enumerate}
\item $ p = 1 $: in this case, the qubit system acts as a single spin $ S = \nspin / 2 $. The operators $ O_{2k-1} $ are all 
proportional to $ S_y $, as shown below. Thus, one variational parameter is sufficient to recover the limit $ l\to\infty $ 
of the NC ansatz. We will numerically show that this corresponds to the exact counterdiabatic potential, up to numerical errors.
		
\item $ p = 2 $: the system exhibits a second-order quantum phase transition, where the minimal gap $ \mingap $ scales as 
$ \mingap \sim \nspin^{-1/3} $. In this case, the CD operator derived within the nested commutator ansatz improves the success 
probability of quantum annealing by increasing the fidelity as a function of the order $ l $. We show that the number of 
NC required grows with the number of qubits $ \nspin $.

\item $ p = 3 $: the system exhibits non degenerate ground state.  It shows  a first-order QPT. The exponent $ p = 3 $ is the smallest 
odd integer for which the $ p $-spin model has this property. Even for this (relatively) simple case, a large number $ l $ of NC in 
Eq.~\eqref{eq:nested-commutators} is needed in order to have a significant improvement in the success probability of quantum 
annealing. Moreover, this number increases with the system size. However, in the following we show that the cyclic ansatz yields an almost perfectly-efficient and size-independent counterdiabatic driving in the 
symmetry subspace with maximum spin. 
\end{enumerate}

Note that the $ p $-spin model Hamiltonian is $ \mathrm{SU}(2) $ invariant, hence states with different total spin $S^2$ belong to 
disconnected subspaces.  If we apply a quantum annealing protocol using the Hamiltonian of Eq.~\eqref{eq:p-spin} as target 
Hamiltonian and $ \hamtf = -\sum_i \sigma_i^x $ as starting Hamiltonian, we can work out the whole procedure within the maximum 
spin subspace, whose dimension is $ \n = \nspin + 1 $. This will allow us to perform numerical simulations for large system sizes.
Moreover, while in the definition of $ S_l(\vec{\alpha}) $ of Eq.~\eqref{eq:action} the traces are evaluated over the $ 2^\nspin $ 
basis states of the full Hilbert space, in this specific case we can define another family of functionals $ \tilde{S}_l(\vec{\alpha}) $, 
in which the traces are restricted to the $ \n = \nspin + 1 $ states of the maximal spin subspace.  In this subspace, $ \tilde{S}_l(\vec{\alpha}) $ 
obeys variational equations analogous to those for $ S_l(\vec{\alpha}) $.
		
We will show that both the nested commutator ansatz and the CA can be studied in the whole Hilbert space or in the maximum spin 
subspace, with different performances. These two approaches are compared in Appendix~\ref{app:whole-vs-maximum}.

	\subsection{Results for $ p = 1 $}
The $p=1$ case is trivial, it is however useful to set the stage for further calculations.	
		The Hamiltonian of the $ p $-spin model for $ p = 1 $ is
		\begin{equation}\label{eq:pspin-p-1}
			\ham_0(\lambda) = -2 (1-\lambda) S_x - 2 \lambda S_z.
		\end{equation}
		It is easy to see that $ O_0 = 2(S_x - S_z) $, while $ O_{2k - 1} \propto S_y $ for all $ k > 0 $. Thus, the ansatz of Eq.~\eqref{eq:nested-commutators} contains only one variational parameter ($ l = 1 $) and is
		\begin{equation}\label{eq:ansatz-p-1}
			A_\lambda^{(1)} = \alpha S_y.
		\end{equation} 
		Details on the minimization of the corresponding quadratic action is given in Appendix~\ref{ansatz}.
		
\begin{figure}[tb]
			\centering
			\includegraphics[width = \linewidth]{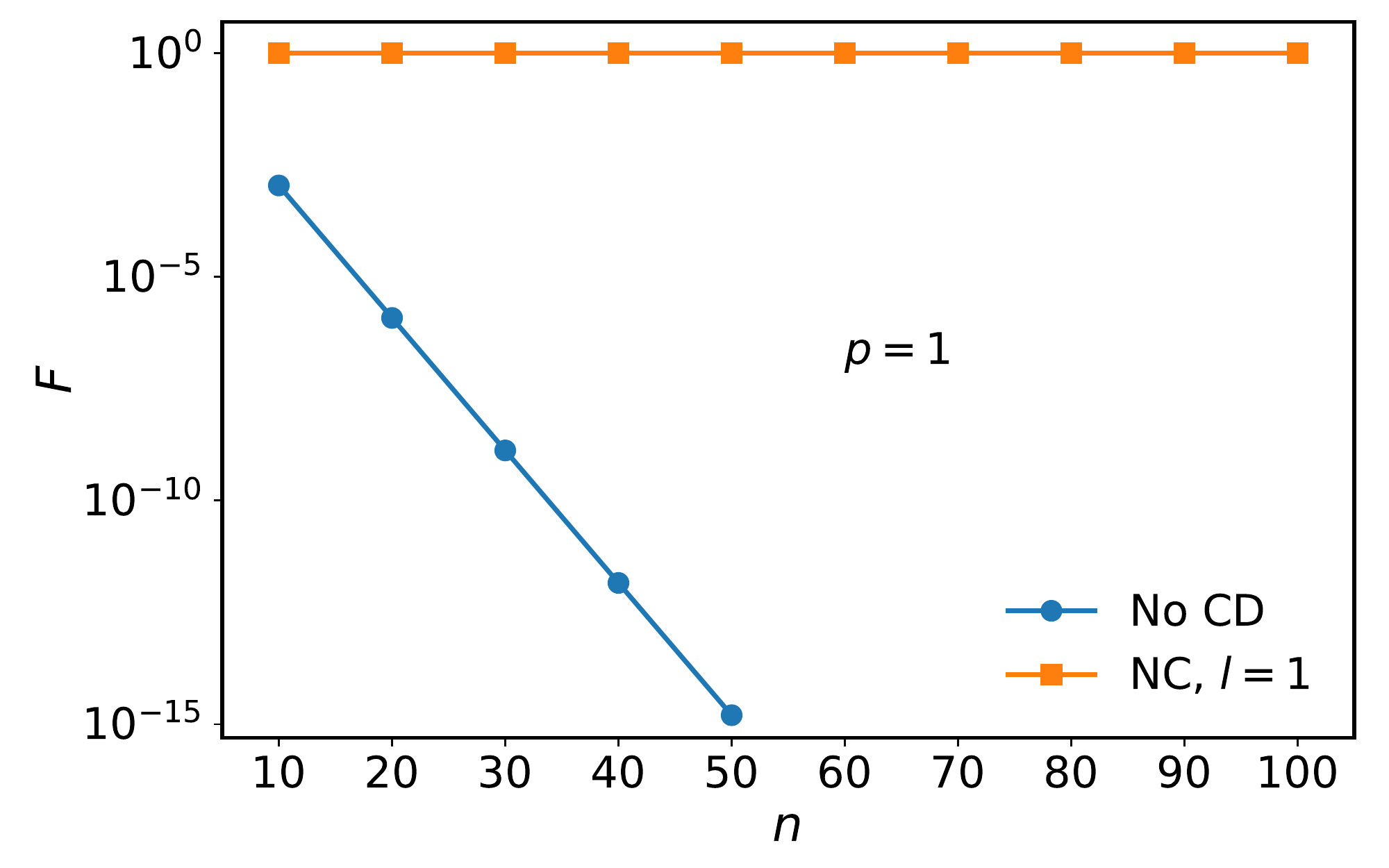}
			\caption{Fidelity $ F $ as a function of the system size, for the bare annealing (blue line with circles) and for the CD driving (orange line with squares), for $ p = 1 $. The ansatz of Eq.~\eqref{eq:ansatz-p-1} yields a size-independent fidelity, up to truncation errors of the numerical integrator.}
			\label{fig:p-1-scaling}
		\end{figure}

		We numerically simulate the dynamics of the $ p $-spin system in the symmetric spin sector, for $ \tf = 1 $, for system sizes ranging from $ \nspin = \text{\numrange{10}{100}} $. The time evolution operator $ U(t) = T_+ \exp(-\iu \int_{0}^{t} \ham(t') dt') $ is approximated discretizing the time interval $ [0, \tf] $ into $ N_t $ steps of $ dt $ and using the following rule:
		\begin{equation}\label{eq:time-evolution}
			U(t) \approx \prod_{i = 1}^{N_t} \eu^{-\iu \ham(t_i + dt / 2) dt}.
		\end{equation} 
		%
		The probability of being in the instantaneous ground state is 
		\begin{equation}
			\pgs(t) = {\lvert \braket{\epsilon_0(t) | U(t) | \psi(t = 0)} \rvert}^2,
		\end{equation}
		and the fidelity is 
		\begin{equation}
			F = \pgs(t = \tf),
		\end{equation}
		 \ie, the probability of being in the ground state at the annealing time $ t = \tf $. 
		 In the absence of the CD term, the fidelity $ F $  is very small in the analyzed cases for this choice of $ \tf $ ($ F \approx \num{1e-3} $ for $ \nspin = 10 $, $ F < \num{1e-15} $ for $ \nspin = 50 $ and above). The scaling of the fidelity as a function of the system size is summarized in Fig.~\ref{fig:p-1-scaling}, up to $ \nspin = 100 $. This clearly shows that the ansatz of Eq.~\eqref{eq:ansatz-p-1} indeed, in this simple case, gives  the exact counterdiabatic potential of Eq.~\eqref{eq:cd-exact}.
		
		
	\subsection{ Results for $p = 2 $}
		
		For $ p = 2 $, there are two degenerate ground states at $ t = \tf $. They both belong to the subspace with maximum spin, \ie, they are the two fully spin polarized states $ \ket{\nspin/2} = \ket{\uparrow\uparrow\cdots\uparrow} $ and $ \ket{-\nspin/2} = \ket{\downarrow\downarrow\cdots\downarrow} $.
		
		In this case, the fidelity is given by $ F = P_{\nspin/2} (\tf) + P_{-\nspin/2} (\tf)$.
		For $ p = 2 $, we simulate the quantum annealing up to a final time  $ \tf = 1 $ and study the scaling of $ F $ as a function of the system size in the maximum spin subspace and for different orders of approximation of the counterdiabatic operator $ A_\lambda^{(l)} $ of Eq.~\eqref{eq:nested-commutators}. 
		
		\begin{figure}[tb]
			\centering
			\includegraphics[width = \linewidth]{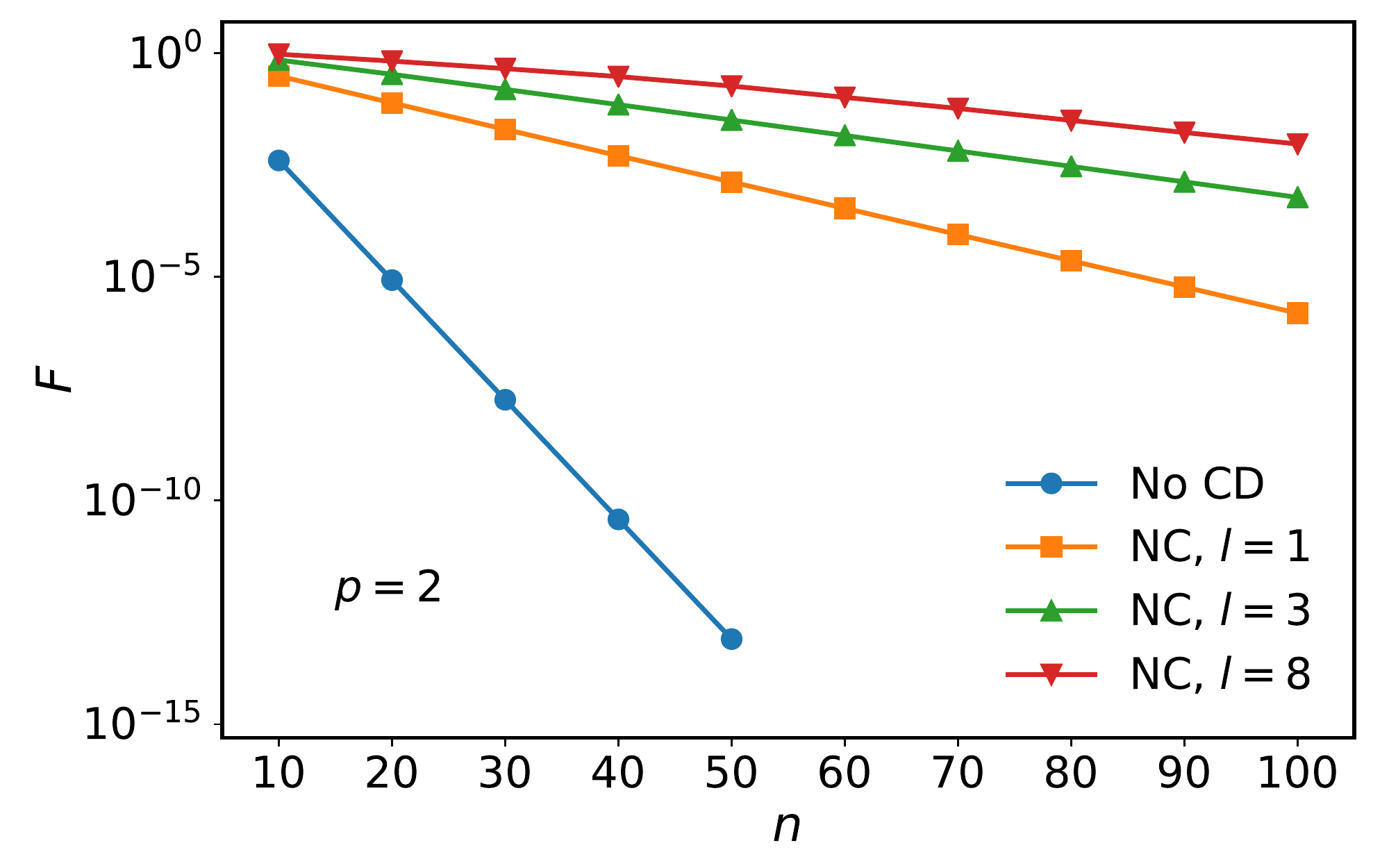}
			\caption{Fidelity  as a function of the system size, for the bare annealing and for the CD ansatz of Eq.~\eqref{eq:nested-commutators} at different orders ($ p = 2 $). Increasing the order of the approximation yields a larger fidelity $ F $.}
			\label{fig:p-2-scaling}
		\end{figure}
		
		In particular, in Fig.~\ref{fig:p-2-scaling} we report the scaling of $ F $ as a function of $ \nspin $, for $ l = \text{\numlist{1;3;8}} $, compared to bare quantum annealing with no CD terms. These results are obtained using a time step of $ dt = \num{1e-3} $. In the bare annealing case, the fidelity rapidly goes to zero as we increase the number of qubits $ \nspin $. This is easily understood, as increasing the number of qubits $ \nspin $ the energy levels become more and more dense and, for fixed $ \tf $,  the dynamics quickly leaves the adiabatic regime. The starting paramagnetic state is metastable for all $ t $ and the $ p $-spin system occupies this state for the whole dynamics.  At the annealing time $ t = \tf $, the system state would be
		\begin{equation}\label{eq:paramagnetic-state}
			\ket{\psi(\tf)} \approx \frac{1}{2^{\nspin/2}} \sum_{w = 0}^\nspin {\binom{\nspin}{w}}^{1/2} \Ket{\frac{\nspin}{2}-w},
		\end{equation}
		where $ \ket{\nspin/2 - w} $ are eigenstates of $ S_z $ with eigenvalues $ \sigma_w = \nspin/2 - w $.
		Only two terms of this sum contribute to the fidelity:
		\begin{equation}\label{Fidelity2}
		F =  {\lvert \braket{-\nspin/2 | \psi(\tf)} \rvert}^2 + {\lvert \braket{\nspin/2 | \psi(\tf)} \rvert}^2 \sim 2^{-\nspin}.
		\end{equation}

		A single variational parameter ($ l = 1 $) yields good improvement for small systems, but eventually the fidelity goes to zero for large $n$.  By increasing the number of variational parameters, the improvement in $F$ can even be of several orders of magnitude for small systems. However, the general trend is that, for large $ \nspin $, this improvement still gives a fidelity  very close to zero and  the proposed variational ansatz seems to be inefficient in the thermodynamic limit.
					
	\subsection{Results for $ p = 3 $}\label{subsec:p-3-symmetry}
		
		For   $ p = 3 $, the ground state is nondegenerate. In  this specific case, we discovered an alternative ansatz, which we named CA, yielding strikingly large fidelities in the symmetric sector, almost independent of the system size. The CA is shown in Eq.~\eqref{eq:sy-ansatz}. In Appendix~\ref{CA_appendix}, we show that for $p=3$ it takes the compact form
\begin{equation}\label{eq:sy-ansatz3}
A_\lambda\api{CA} = \alpha_1 S_y +  \alpha_3 S_y^3 + \alpha'  (S_x S_y S_z + \hc),
\end{equation}  
having only three variational parameters.
		
We start this section showing the scaling of the fidelity as a function of the system size. Due to the first-order QPT, we expect that the fidelity scales as
\begin{equation}\label{fit}
  F = \phi \eu^{-\gamma \nspin }.
\end{equation}
We ask whether the CD driving can change this scaling law or not.

\begin{figure}[tb]
	\centering
	\includegraphics[width = \linewidth]{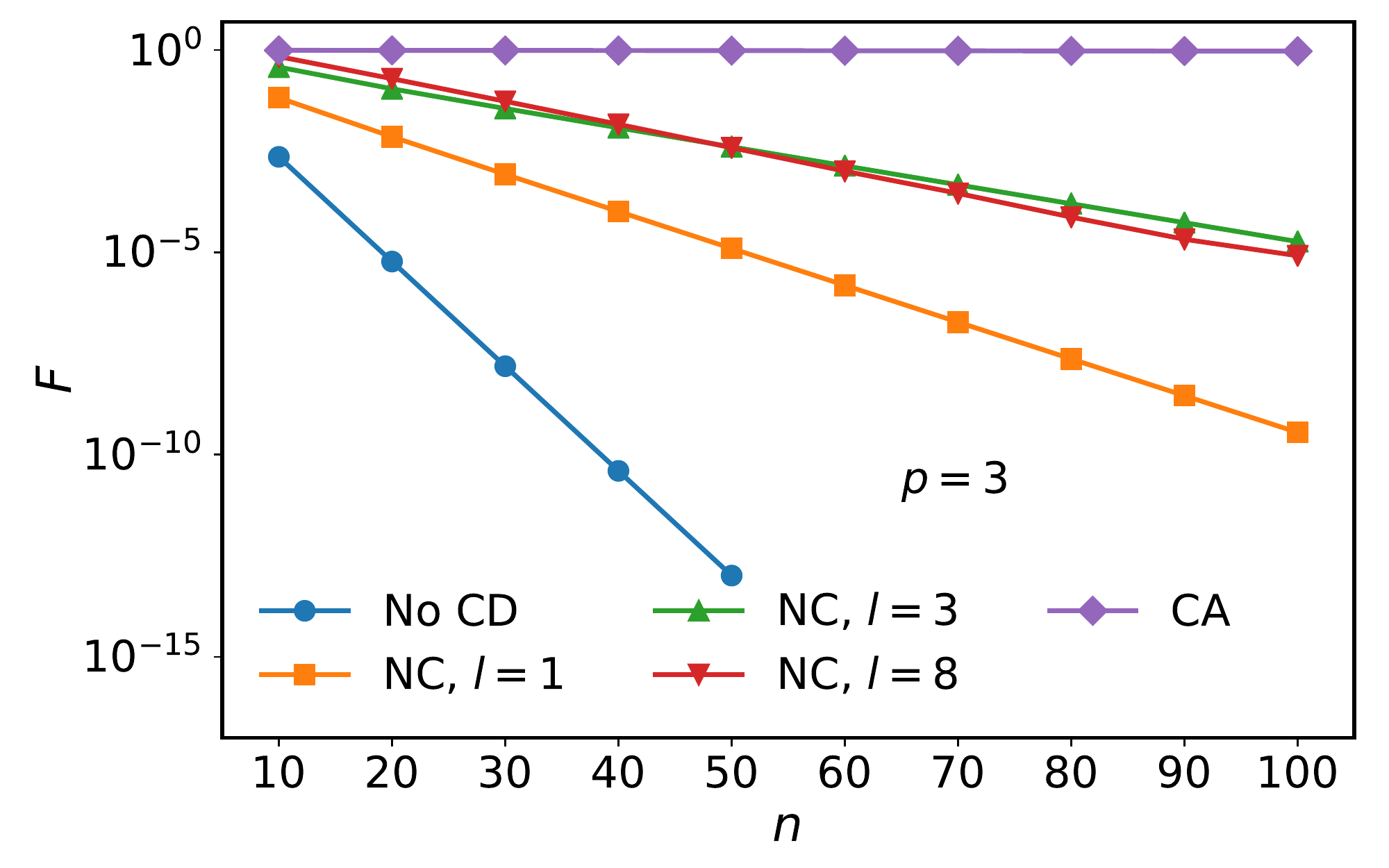}
	\caption{Fidelity $ F $ as a function of the system size, for the bare annealing and for the CA and NC ansatz at different orders ($ p = 3 $).}
	\label{fig:p-3-scaling}
\end{figure}

In Fig.~\ref{fig:p-3-scaling}, we show the fidelity $ F $ as a function of the system size for $ \tf = 1 $ and we compare the bare quantum annealing with the nested commutator ansatz ($ l = \text{\numlist{1;3;8}} $) and the CA. We perform a fit of the results conjecturing the exponential behavior of Eq.~\eqref{fit} even in the presence of CD.
The coefficients $\phi$ and $\gamma$ are summarized in Table~\ref{table1}, comparing the standard quantum annealing, the NC ansatz (for several orders $l$), and the CA.

\setlength{\tabcolsep}{1em}
\begin{table}
	\caption{Table of coefficients for the exponential fit in Eq.~\eqref{fit}.}
	\label{table1}
	\begin{tabular}{lcc}
		\toprule
					& {$\phi$} & {$\gamma$}  \\ 
		\midrule
		No CD   	& \num{0.903}  & \num{5.96e-1}   \\  
		NC, $l=1$	& \num{0.492}  & \num{2.11e-1}   \\ 
		NC, $l=3$	& \num{0.999}  & \num{1.09e-1} 	 \\ 
		NC, $l=8$	& \num{2.438}  & \num{1.29e-1}	 \\ 
		CA			& \num{0.990}  & \num{5.54e-4}   \\
		\bottomrule
	\end{tabular} 
\end{table}

%
Note that the exponent $ \gamma $ in the CA is three orders of magnitude smaller than both the unitary case and the nested commutator ansatz. Moreover, in the latter case, we observe that the fidelity grows with increasing order $ l $ only for small system sizes, whereas for larger systems the fidelity shows a maximum as a function of $ l $ and then decreases. In fact, the exponent $ \gamma $ for $ l = 8 $ is larger than that for $ l = 3 $. 

To summarize, in the presence of a CD the scaling of $ F $ with $ \nspin $ remains exponential, but the coefficient $ \gamma $ is reduced with respect to the bare case. Moreover, the CA yields an almost constant fidelity up to $ \nspin = 100 $, and a large one ($ F > 1/2 $) up to $ \nspin = 1000 $, providing a robust mechanism to counteract the exponentially vanishing spectral gap for macroscopic systems.

	\begin{figure*}[t]
		\centering
		\includegraphics[width = 0.49\linewidth]{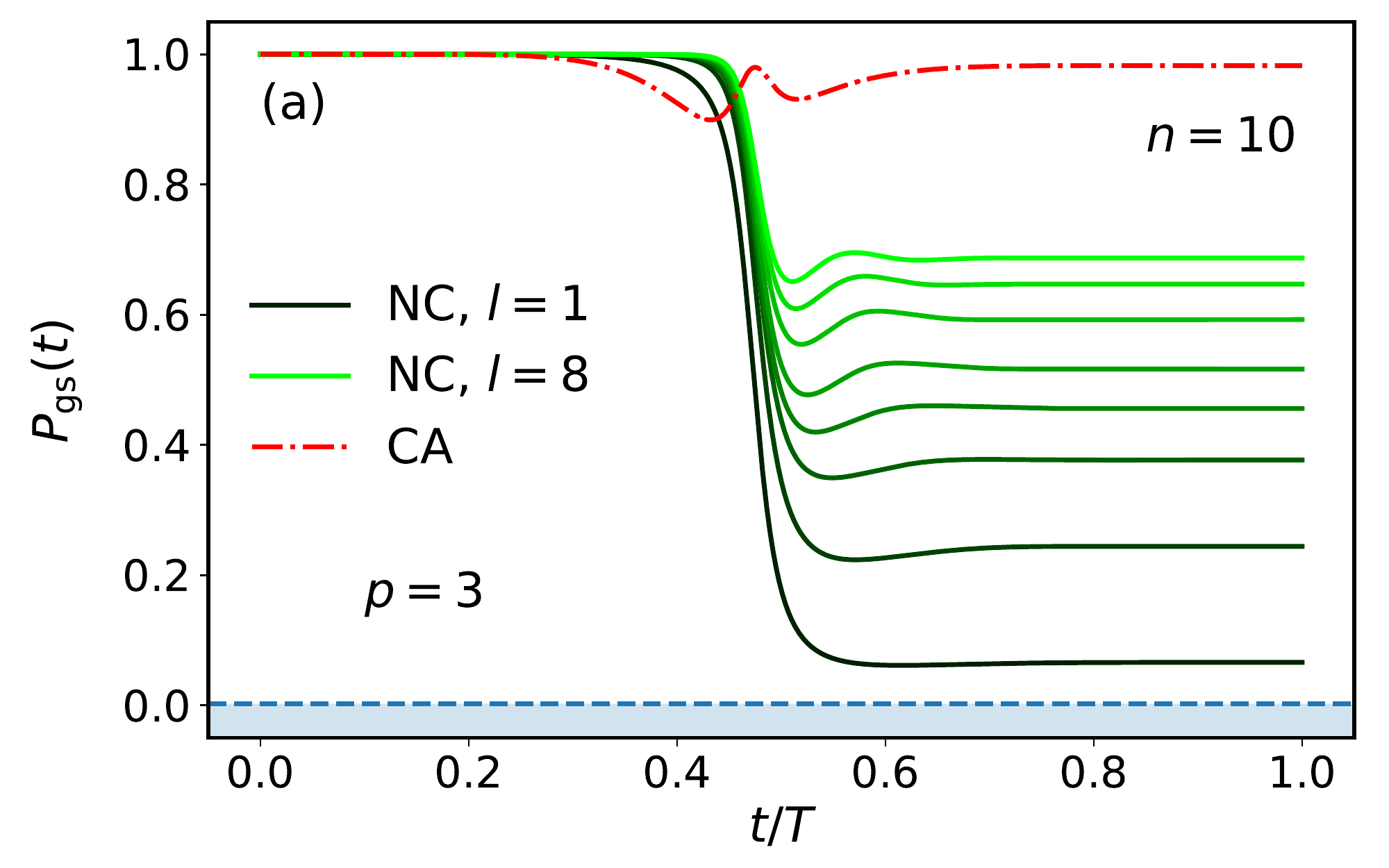}
		\includegraphics[width = 0.49\linewidth]{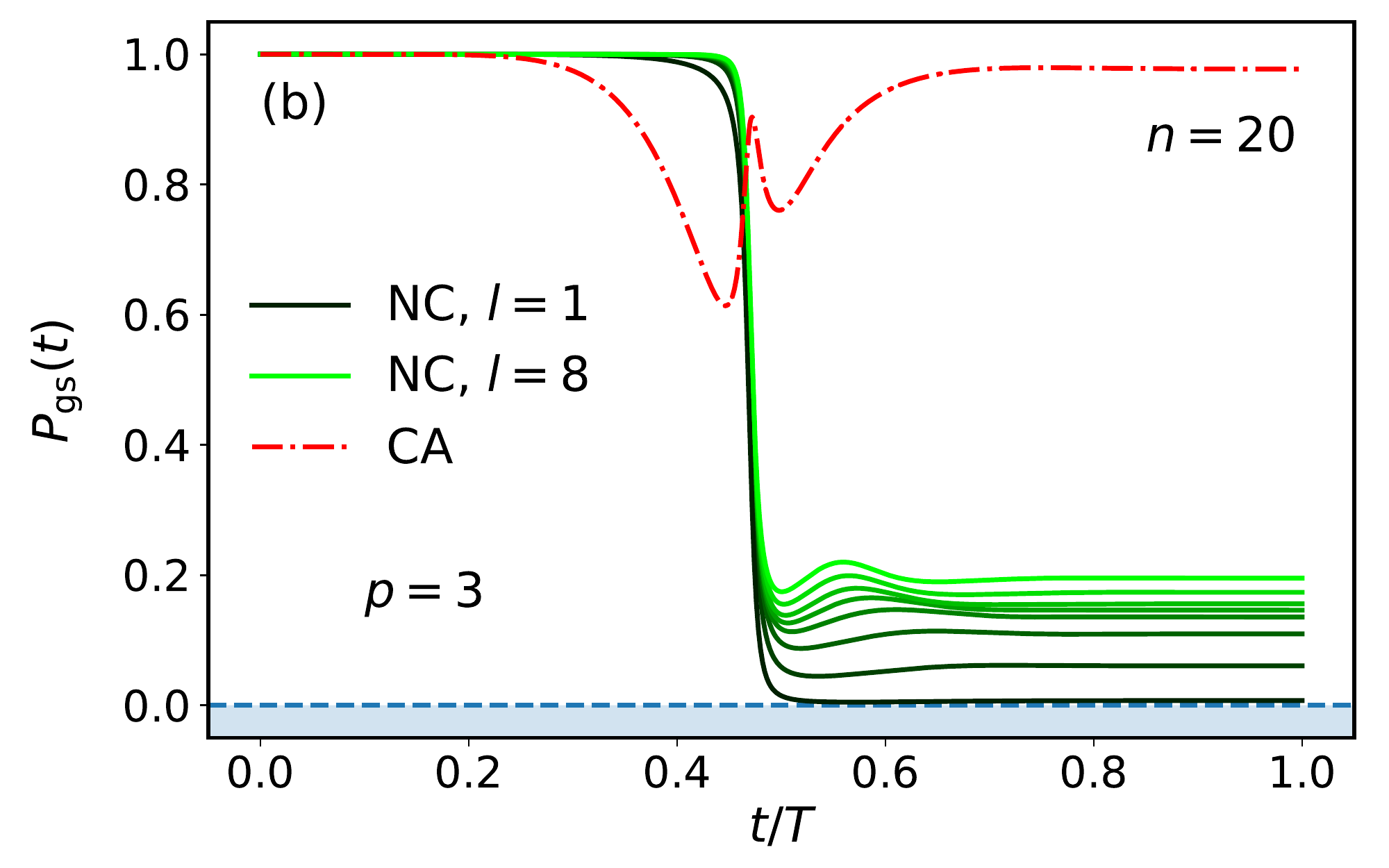}
		\caption{Time dependence of the ground state probability $ \pgs $, for $ p = 3 $, $ \tf = 1 $, and (a) $ \nspin = 10 $ and (b) $ \nspin = 20 $. The blue dashed line represents the fidelity of the bare evolution without CD driving. The darkest line is the ground state probability when the NC ansatz of order $ l = 1 $ is used. Analogously, the lightest line refers to $ l = 8 $ and all lines in between correspond to increasing orders of the NC ansatz [Eq.~\eqref{eq:nested-commutators}]. The red dotdashed line refers to the CA.}
		\label{fig:p-3-dynamics}
	\end{figure*}

		
		In Fig.~\ref{fig:p-3-dynamics}, we show the time evolution of the ground state probability $ \pgs(t) $ for $ \tf = 1 $, for $ \nspin = 10 $ [panel (a)] and $ \nspin = 20 $ [panel (b)] in the maximal spin subspace. The blue dashed line indicates the fidelity of bare quantum annealing with no CD. The lightest green line corresponds to a CD dynamics with $ l = 8 $. Darker lines are for all orders starting from $ l = 1 $ (the lighter is the line, the larger is $ l $). The red dotdashed line represents the CA. Fig.~\ref{fig:p-3-dynamics} clearly shows that the nested commutator ansatz of Eq.~\eqref{eq:nested-commutators} can improve the fidelity of quantum annealing for $ p = 3 $. However, the CA yields an even larger fidelity. In comparison, a similar fidelity could be reached only  going beyond order $ l = 8 $. However, increasing the number of NC the improvement in the fidelity is gradually smaller, and we guess that in order to achieve results similar to the one obtained using the CA an unpractical large number of nested commutator would be required. Moreover, as $ \nspin $ grows, more and more variational parameters are required to achieve a similar level of fidelity, whereas  the CA requires only three variational parameters. Hence, in this particular case, the CA is extremely efficient and outperforms the other known approximation schemes.
This is even more evident increasing the number of qubits.	 For instance, Fig.~\ref{fig:p-3-dynamics}(b) shows the same results for $ \nspin = 20 $. Here, the fidelity for $ l = 8 $ is significantly smaller than the previous case ($ F \approx 0.20 $ versus $ F \approx 0.68 $). By contrast, the fidelity of the CA is almost unchanged with respect to $ \nspin = 10 $, while the ground state probability at intermediate times is still affected by the system size. 
		

\section{Finite range and random instances}
\label{non-pspin}
	
		The efficiency of the CA could depend on the peculiarities of the $ p $-spin model, \ie, spin symmetry and infinite-range interactions. 
However, it is difficult to prove this statement theoretically, therefore in the following we will try to limit the range of the interactions and to break spin symmetry to gain some insights into this problem.
In the absence of spin symmetry, we need to extend the analysis presented in Sec.~\ref{subsec:p-3-symmetry} to the whole Hilbert space. Of course, this process is exponentially more demanding, since now we have to consider all the $ 2^\nspin $ computational basis states. Therefore, in this section, we will limit our analysis to small systems, \ie, $ \nspin = \text{\numrange{3}{8}} $.
		
The total Hilbert space can be decomposed as the direct sum of the  eigenspaces of the maximum total  spin operator, corresponding to $ S = \nspin/2 $, and of its orthogonal:
		\begin{equation}\label{eq:hilbert-spaces}
			\mathcal{H} = \mathcal{H}_{\nspin/2}  \oplus \mathcal{H}^\perp_{\nspin/2}.
		\end{equation}
		In principle, all traces in Eq.~\eqref{eq:action-quadratic} have to be evaluated over $ \mathcal{H} $. However, we have already discussed that the variational procedure can be easily restricted to the interesting subspace. Using a parameter $ 0\le \eta \le 1 $, we can choose whether to evaluate the traces in the whole Hilbert space or rather in the symmetric subspace. For any operator $ O $, we replace
		\begin{equation}
			\Tr (O) \longrightarrow (1-\eta) \Tr(O)_{\nspin/2} + \eta \Tr(O)_{\nspin/2}^\perp.
		\end{equation}
		The case $ \eta = 0 $ is analogous to Section~\ref{subsec:p-3-symmetry}. Minimizing in the whole Hilbert space corresponds to choosing $ \eta = 1/2 $. The comparison between the two approaches is discussed in Appendix~\ref{app:whole-vs-maximum}. Finally, $ \eta = 1 $ is the minimization in the orthogonal subspace $ \mathcal{H}^\perp_{\nspin/2} $. Here  focus our attention to the cases in which  $ \eta = 0 $ and $ \eta = 1/2 $.
		
		\subsection{Finite-range $ p $-spin model}
		
		To highlight the infinite-range nature of the $ p $-spin model, it is more convenient to rewrite its Hamiltonian [Eq.~\eqref{eq:p-spin}] in the following form:
		\begin{equation}\label{eq:p-spin-expanded}
			\hampr = -\frac{1}{\nspin^{p - 1}} \sum_{i_1, \dots, i_p} J \sigma_{i_1}^z \dots \sigma_{i_p}^z,
		\end{equation}
		where $ i_j = 1, \dots, \nspin $ for all $ j $. As $ \lbrace \sigma_i^z, \sigma_j^z \rbrace = 2 \delta_{i, j} $, the Hamiltonian in Eq.~\eqref{eq:p-spin-expanded} is a polynomial function of order $ p $ of Pauli operators, containing terms of orders $  P = p, p - 2, \dots, 1 $ (odd $ p $). Each term represents an infinite-range $ P $-body interaction between qubits, with uniform coupling constant $ J $.
		
		A possible way for turning this infinite-range $ p $-spin model into a finite-range model is by weighting $ J $ with the  ``distance'' between the qubits involved in the $ p $-body term. This can be easily understood by considering the simple case $ p = 2 $, where the Hamiltonian would be
		\begin{equation}
			\hampr^{p = 2} = \text{const} - \frac{2}{\nspin} \sum_{i, j} J \sigma_i^z \sigma_j^z.
		\end{equation}
		In this case, we can replace $ J \to J_{i,j} = J / {\lvert i - j \rvert}^\nu $ and build a finite-range version of the $ p $-spin model, where the exponent $ \nu $ determines how punctual the interactions between the qubits are: $ \nu = 0 $ is the infinite-range model and $ \nu \to \infty $ represents nearest-neighbor interacting qubits.
		
		The same reasoning holds for any value of $ p $. In particular, we can always replace $ J $ by $ J_{i_1, \dots, i_p} = J/\dist(i_1, \dots, i_p)^\nu $. Here, we propose to consider the following form for the distance function:
		\begin{equation}\label{eq:dist}
			\dist(i_1, \dots, i_p) = 
			\begin{cases}
				\sum_{j, k > j} \lvert i_k - i_j \rvert / Z & \text{if $ i_j $'s are all distinct;}\\
				1 &\text{otherwise.}
			\end{cases}
		\end{equation}
		The parameter $ Z = (p^3 - p)/ 6 $ is a normalization factor chosen so that $ \dist(1, 2, \dots, p) = 1 $. For $ p = 3 $, that normalization factor is $ Z = 4 $. We also note that this choice of the distance function does not allow finite-range models for $ \nspin = 3 $ with $ p = 3 $, as in that case $ J_{i_1, i_2, i_3} = 1 $ for all combinations of indices. Of course, for all other values of $ \nspin $, this procedure breaks the spin symmetry of the $ p $-spin model, therefore we will work in the whole Hilbert space and consider $ \eta = 1/2 $.
		
		\begin{figure*}[t]
			\centering
			\includegraphics[width = 0.49\linewidth]{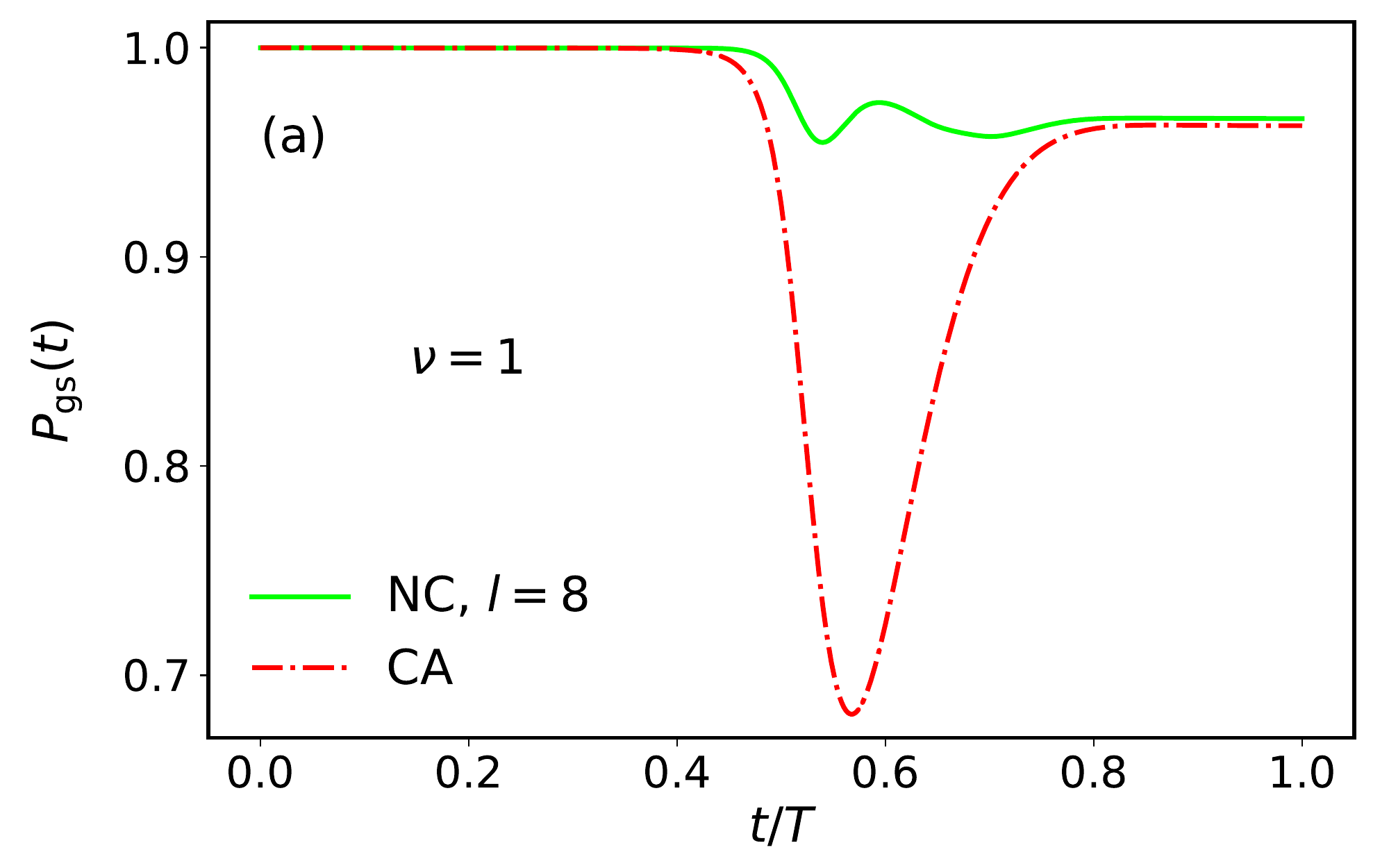}
			\includegraphics[width = 0.49\linewidth]{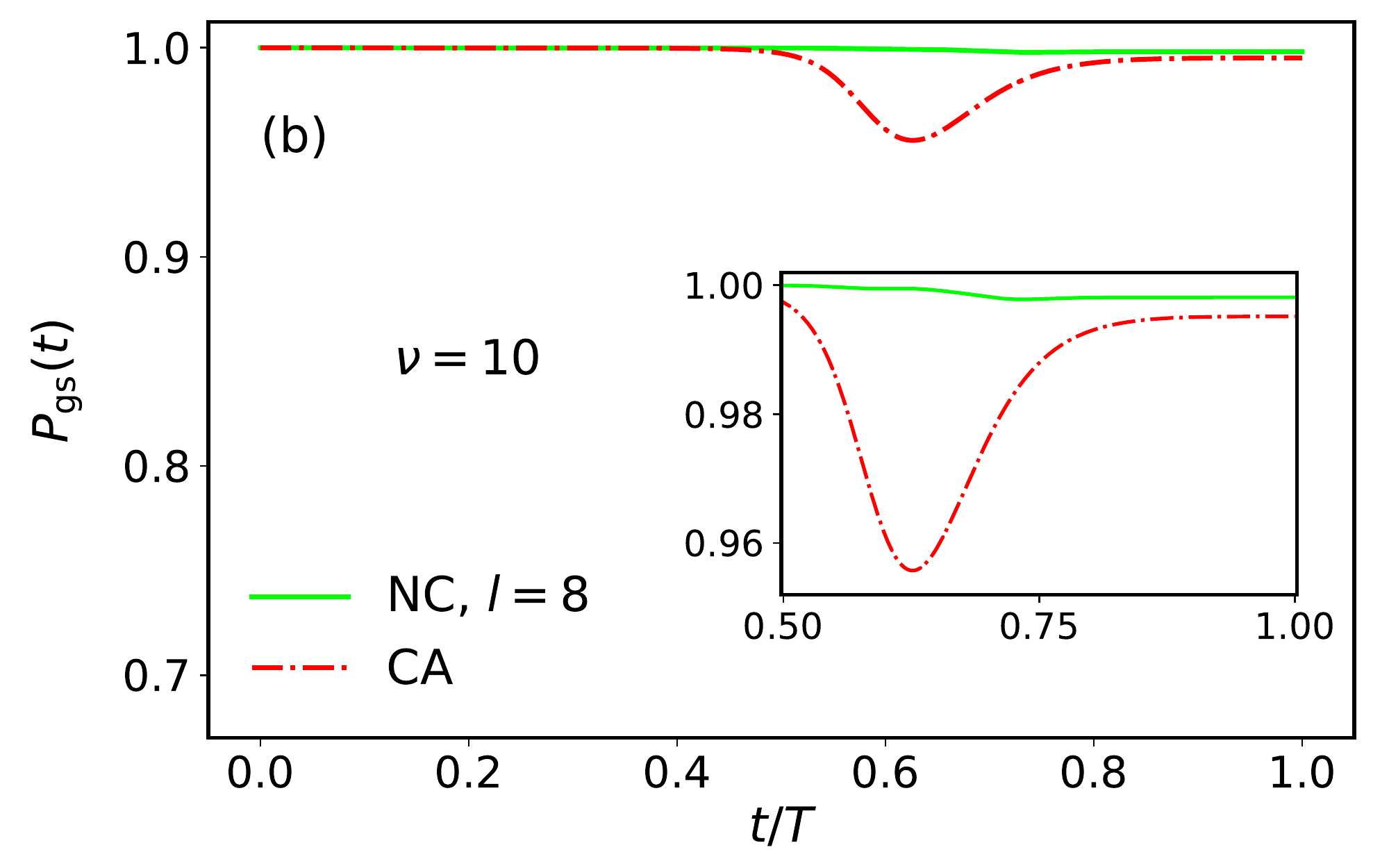}
			\caption{Ground state probability $ \pgs(t) $ as a function of time, for $ \nspin = 8 $ and $ p = 3 $. The annealing time is $ \tf = 1 $. The minimization occurs in the whole Hilbert space ($ \eta = 1/2 $). Panel (a) is for $ \nu = 1 $ and panel (b) is for $ \nu = 10 $ in Eq.~\eqref{eq:dist}. In both panels, the green solid line refers to the NC ansatz ($ l = 8 $), while the red dotdashed line refers to the CA. The inset in panel (b) zooms in on the region $ t/\tf \in [0.5, 1] $.}
			\label{fig:n-8-p-3-nosymm-ranges}
		\end{figure*}
		
		In Fig.~\ref{fig:n-8-p-3-nosymm-ranges}, we show the ground state probability $ \pgs(t) $ as a function of time, for $ \nspin = 8 $ and $ p = 3 $. The green solid line corresponds to NC while the red dotdashed line corresponds to the CA. The left-hand (right-hand) panel refers to $ \nu = 1 $ ($ \nu = 10 $). This figure can be compared with the right-hand panel of Fig.~\ref{fig:n-8-p-3-symm-nosymm}, corresponding to $ \nu = 0 $. Moving from the infinite-range model to the finite-range one, we note that the efficiency of both ans\"atze, NC and CA, is improved, as both curves are pushed upwards. In particular, for $ \nu = 10 $, the fidelity $ F $ in the CA case is $ F \approx 0.995 $, comparable with the nested commutator ansatz with $ l = 4 $.
		Can we conclude that the reason why the CA works so well cannot be the fact that the model is infinite-range as it works even better without this feature? It is difficult to compare to the $ p $-spin model as 1) we are not working in the symmetric subspace and 2) the two models have different spectra (different gap, different time of minimal gap, see the unitary dynamics), however the results of this section motivated us to go even deeper and to analyze different (random) instances and check our ansatz also in that case.
		
		\subsection{Random $ p $-spin model}
		
		Previously, we showed that the CA can be efficient also for models featuring finite-range interactions, as in that case we can even get larger fidelities than those of the infinite-range ferromagnetic $ p $-spin model. In this section, we will address another question. Starting from the Hamiltonian of Eq.~\eqref{eq:p-spin-expanded}, here we randomly suppress some of the coupling constants $ J $ with a certain probability. In this way, we can build a family of infinite-range models, where the full-connectivity of the $ p $-spin model is progressively lost. 
		
		The resulting Hamiltonian is identical to that in Eq.~\eqref{eq:p-spin-expanded}, but the coupling constant $ J $ is replaced by a random variable $ K $ satisfying
		\begin{equation}
			K = 
			\begin{cases}
				J & \text{with probability $ P_J $;}\\
				0 & \text{with probability $ 1 - P_J $.}
			\end{cases}
		\end{equation}
		This model is the usual infinite-range ferromagnetic $ p $-spin model when $ P _J = 1 $. For any $ P_J \ne 0, 1 $, this model breaks the spin symmetry and we have to work in the whole Hilbert space, with $ \eta = 1/2 $.	
		
		\begin{figure}
			\centering
			\includegraphics[width = \linewidth]{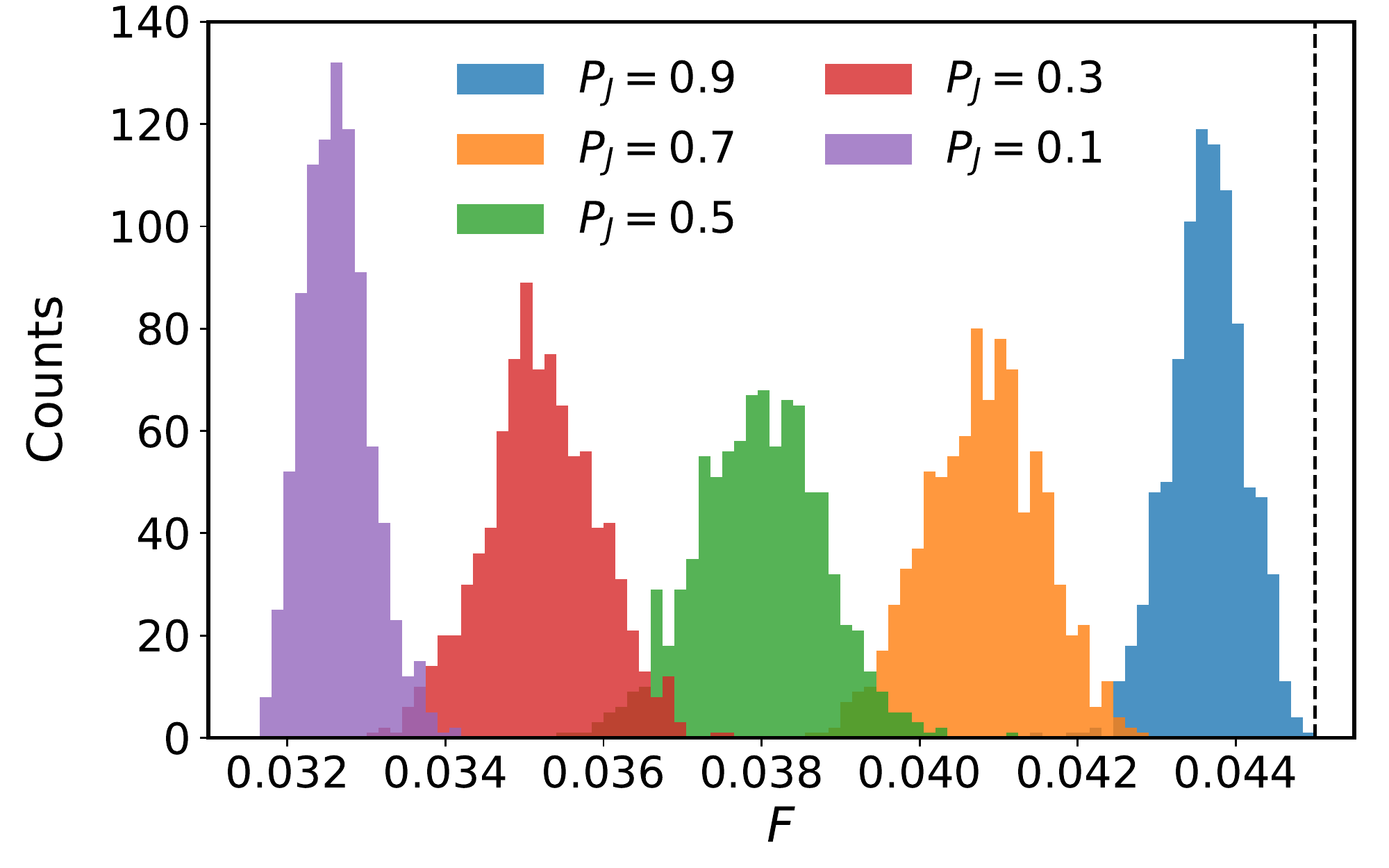}
			\caption{Fidelity distributions for the random $ p $-spin model for several values of the probability $ P_J $ of having nonzero coupling constants, for $ \nspin = 5 $ and $ p = 3 $. The annealing time is $ \tf = 1 $ and there are no counterdiabatic terms. The vertical dashed line indicates the fidelity for $ P_J = 1 $.}
			\label{fig:histogram-n-5-p-3-no-cd}
		\end{figure}
		\begin{figure*}[t]
			\centering
			\includegraphics[width = 0.49\linewidth]{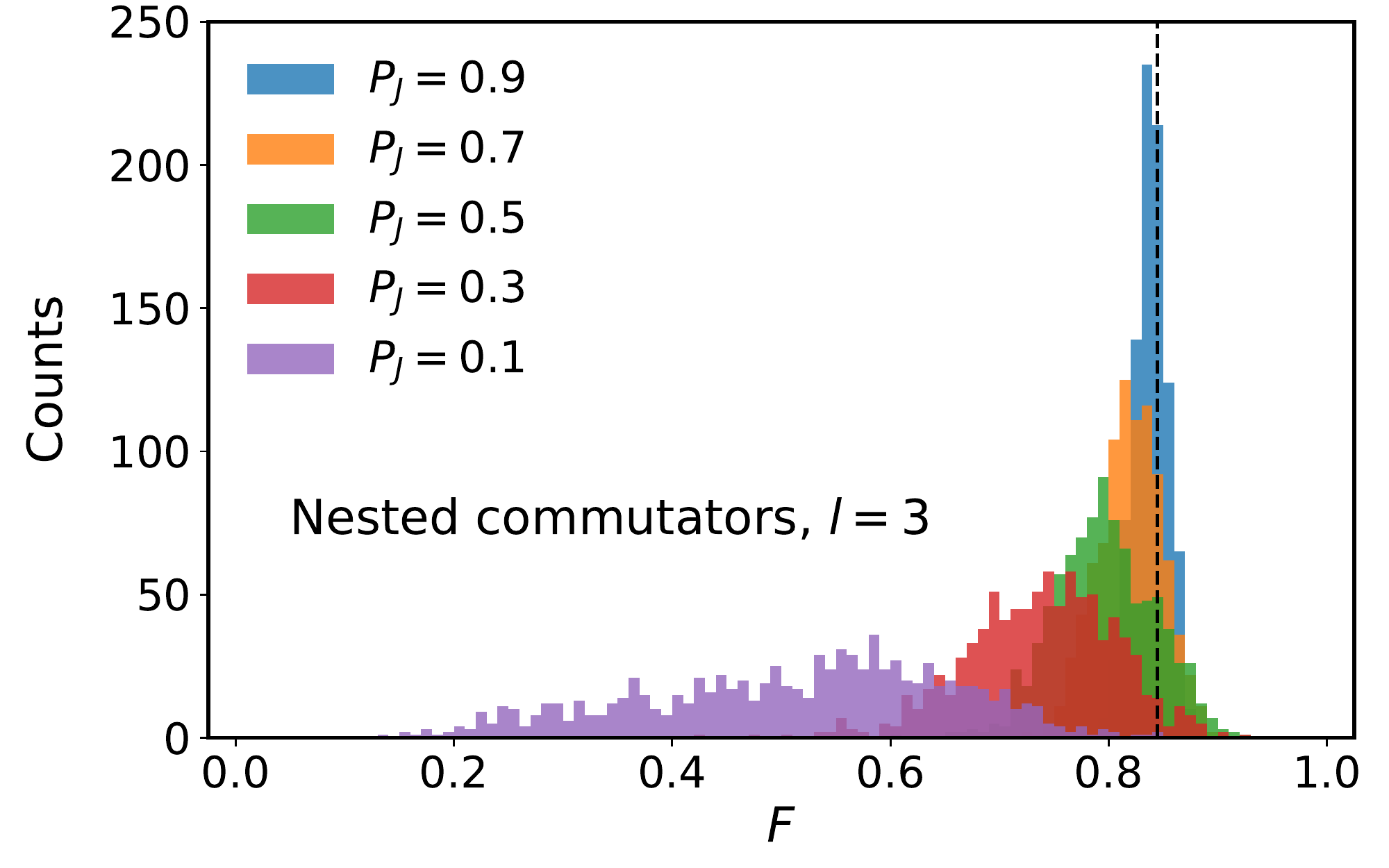}
			\includegraphics[width = 0.49\linewidth]{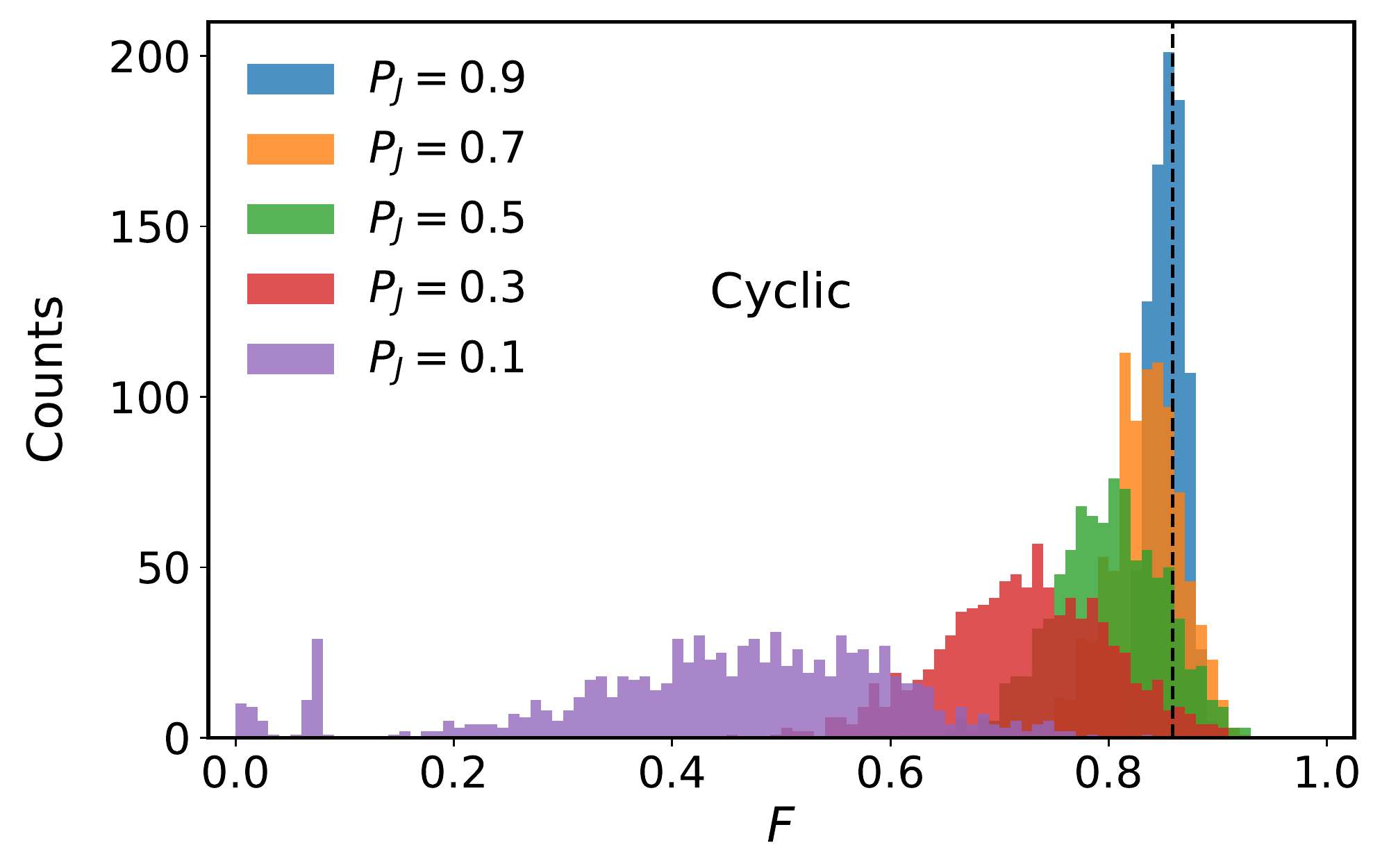}
			\caption{Fidelity distributions for the random $ p $-spin model, in the presence of counterdiabatic driving. The left-hand panel refers to the nested commutator ansatz with $ l = 3 $. In the right-hand panel, we show data for the CA. In both panels, the black dashed line indicates the fidelity for $ P_J = 1 $.}
			\label{fig:histogram-n-5-p-3-cd}
		\end{figure*}
		
		For several choices of $ P_J $, we performed dynamical simulations for $ M $ randomly generated instances of this infinite-range random $ p $-spin model and measured the fidelity, both with and without counterdiabatic terms. We focus here on the case $ \nspin = 5 $ and $ p = 3 $, however we obtained qualitatively similar results also for larger system sizes.	
				
		In Fig.~\ref{fig:histogram-n-5-p-3-no-cd}, we show the fidelity distributions for $ P_J = \text{\numlist{0.1;0.3;0.5;0.7;0.9}} $. We divided the fidelity interval $ [0.30, 0.45] $ into $ N\ped{b} = 100 $ bins and counted the occurrences over $ M = 900 $ repetitions of the dynamics. 

The peaks of the distributions are equally spaced, which  implies that the mean fidelity $ \langle F \rangle $ linearly depends on $ P_J $. 
According to the Landau-Zener formula, in a two-level approximation around the avoided crossing the mean fidelity would be
		\begin{equation}
		\langle F \rangle \approx 1 - \eu^{2\pi \langle\mingap\rangle^2 \tf} \approx 2\pi \langle\mingap\rangle^2 \tf.
		\end{equation} 
As a consequence $\langle F \rangle \approx \langle\mingap\rangle^2 \propto P_J $ which implies that  $ \langle\mingap\rangle \sim P_J^{1/2} $. 
			
		As $ P_J < 1 $, all randomly generated instances have smaller gaps than the infinite-range ferromagnetic $ p $-spin model with $ P_J = 1 $, therefore the corresponding fidelity is always smaller than that of the original model in the absence of counterdiabatic terms. This is shown in Fig.~\ref{fig:histogram-n-5-p-3-no-cd}, using a black dashed line to highlight the fidelity for $ P_J = 1 $.
	
		In Fig.~\ref{fig:histogram-n-5-p-3-cd}, we show the same fidelity distributions in the presence of counterdiabatic driving. The left-hand panel is for the NC ansatz of order $ l = 3 $. The right-hand panel is for the CA. Here, we consider $ N\ped{b} = 100 $ bins for the fidelity interval $ [0, 1] $. Except for a few minor differences, such as the presence of outliers around $ F = 0 $ for $ P_J = 0.1 $, the two plots look similar. However, we note that the mean fidelity in the cyclic case is smaller than that in the NC case for $ P_J \le 1/2 $, while the opposite is true for $ P_J > 1/2 $. In both cases, the presence of counterdiabatic driving allows for a significantly larger fidelity ($ \sim \text{\numrange[range-phrase=--]{15}{20}} $ times), compared to the case with no CD driving. Increasing the order $ l $ of the NC ansatz, the mean values of all distributions move towards $ F = 1 $. 
		Moreover, all distributions become narrower and acquire nonzero skewness. 
		
		As opposed to the case with no CD driving, here there are some instances showing larger fidelity than that for $ P_J = 1 $. This is highlighted in Fig.~\ref{fig:histogram-n-5-p-3-cd}, 
		where the black dashed line indicate the fidelity of the case $ P_J = 1 $. This evidence confirms that the efficiency of counterdiabatic driving does not entirely depend on the spectral properties of the analyzed model. In fact, even if the average minimal gap is smaller than that for $ P_J = 1 $, there are instances of the $ P_J < 1 $ case where quantum annealing with CD is more efficient than the case $ P_J = 1 $ with CD. 
				
\section{Conclusions}
\label{conclusions}
To summarize, we have applied the  variational approach (developed in Ref.~\cite{SelsPolkovnikov}) to find an approximate CD potential for the quantum annealing of the $ p $-spin model. This approach is very promising~\cite{KOLODRUBETZ20171} and has been recently applied to other kinds of model Hamiltonians with $p$-body interactions~\cite{Hartmann_2019}. Our results confirm these expectations.
The $ p $-spin model possesses a spin symmetry, hence we first restricted our approach to the maximum spin subspace, where the starting and final states lay.  We used two different ans\"atze for the variational potential: one is based on the NC hypothesis~\cite{FloquetManyBody}, and another based on a cyclic ansatz [see Eq.~\eqref{eq:sy-ansatz}]. We focused our attention on the case $ p = 3 $, which is the first odd integer showing a first-order quantum phase transition with exponentially small gap closure in the thermodynamic limit.  In this case, both ans\"atze give substantial improvement in the fidelity with respect to the bare dynamics (\ie, standard quantum annealing). However, the CA seems to be much more efficient, in particular when the number of qubits grows.
For both ans\"atze, we can also perform the variational optimization in the whole Hilbert space. However, in this case, the efficiency of the CA  is reduced, as discussed in Appendix~\ref{app:whole-vs-maximum}. This is likely due to the fact that the corresponding optimal counterdiabatic potential also addresses diabatic transitions between pairs of energy eigenstates that are already uncoupled by the spin symmetry. This redundancy harms the performances of the CA in the relevant symmetry subspace, where the dynamics occurs. By contrast, restricting the traces to the subspace with maximal spin, we can readily implement the spin symmetry of the $ p $-spin model and obtain better performances.  We have also shown new results concerning modified model Hamiltonians accounting for short-range interactions and random instances. 
These results may be relevant for the future implementation of experimentally viable counterdiabatic operators on the available quantum processors.

\acknowledgments
The authors acknowledge the CINECA Award under the ISCRA initiative (project QA-MCWF) for the availability of high-performance computing resources and support.

\appendix
\section{Variational ansatz}
\label{ansatz}

A possible variational strategy consists in restricting the optimization to physically realistic (local) operators. For instance, the authors of Ref.~\cite{SelsPolkovnikov} prove that the following form,
		\begin{equation}
			A_\lambda^* = \sum_j \alpha_j \sigma_j^y,
		\end{equation}
		where $ \alpha_j $ are variational parameters and $ \sigma_j^k $, $ k = x, y, z $ are Pauli matrices, can successfully approximate the true counterdiabatic potential of the 1D quantum Ising model with both transverse and longitudinal biases. Contextually, they also show that this ansatz is no longer valid when either transverse or longitudinal biases (or both) are zero. The effectiveness of the procedure can be improved systematically by including higher-order local terms such as $ \sigma_j^y \sigma_{j + 1}^z + \sigma_j^y \sigma_{j + 1}^x + \hc $ in the original ansatz. The exact counterdiabatic driving breaks time reversal, thus an odd number of $ \sigma^y $ operators is required at all orders. However, in this approach, it is \textit{a priori} unknown which operators have to be included to have a sufficiently accurate description of the counterdiabatic potential. 
		
		In Ref.~\cite{FloquetManyBody}, the authors proposed the following approximate gauge potential:
		\begin{align}\label{eq:nested-commutators_app}
			A_\lambda^{(l)} &= \iu \hslash \sum_{k=1}^{l} \alpha_k [\underbrace{\ham_0, [\ham_0, \dots [\ham_0}_{2k-1}, \partial_\lambda \ham_0]]] \notag \\
			&= \iu\hslash \sum_{k=1}^{l} \alpha_k O_{2k-1}.
		\end{align}
		For ease of notation, we have defined
		\begin{equation}
			\begin{cases}
				O_k = \comm{\ham_0}{O_{k-1}}, & \text{$ k \ge 1 $}; \\
				O_0 = \partial_\lambda \ham_0.
			\end{cases}
		\end{equation}
		Now the operator $ G_\lambda $ of Eq.~\eqref{eq:operator-G} reads
		\begin{equation}
			G_\lambda^{(l)} = \partial_\lambda \ham_0 + \frac{\iu}{\hslash} \comm{A_\lambda^{(l)}}{\ham_0} = O_0 + \sum_{k = 1}^l \alpha_k O_{2k},
		\end{equation}
		thus the corresponding action is
		\begin{align}
			S_l(\vec{\alpha}) &= \Tr\bigl[O_0^2\bigr] + 2\sum_{k = 1}^l \alpha_k \Tr\bigl[O_0 O_{2k}\bigr] \notag\\
			&\quad+ \sum_{j, k = 1}^l \alpha_j \alpha_k \Tr\bigl[O_{2j} O_{2k}\bigr],
		\end{align}
		we can define $ \vec{\alpha} = (\alpha_1, \dots, \alpha_l) $ and recast  $ S_l $ as a quadratic polynomial
		\begin{gather}\label{eq:action-quadratic}
			S_l(\vec{\alpha}) = A + 2 \vec{B} \cdot \vec{\alpha} + \vec{\alpha}^T \cdot \underline{C} \cdot \vec{\alpha},\\
			A = \Tr\bigl[O_0^2\bigr], \\
			B_i = \Tr\bigl[O_0 O_{2i}\bigr], \\
			C_{ij} = \Tr\bigl[O_{2i} O_{2j}\bigr].
		\end{gather}is diagonal
		We can introduce the matrix $ U $ that diagonalizes $\underline{C}$, \ie, $ \underline{D} = U^T \underline{C} U $. Then, $ S_l $ is rewritten as 
		\begin{equation}
			S_l(\vec{\alpha}') = A + \sum_{k = 1}^{l} \left(2 B'_k \alpha'_k + D_{kk} {\alpha'_k}^2\right),
		\end{equation}
		with $ \vec{\alpha}' = U^T \vec{\alpha} $ and $ \vec{B}' = U^T \vec{B} $. The stationary point is
		\begin{equation}
			\frac{\partial S_l}{\partial \alpha'_k} = 2 B'_k + 2 D_{kk} \alpha'_k = 0 \Longrightarrow \alpha'_k = -\frac{B'_k}{D_{kk}}. 
		\end{equation}
		
		The exact CD driving is recovered in the limit $ l \to \infty $. In fact, the matrix elements of $ A_\lambda^{(l)} $ in the energy basis are
		\begin{equation}
			\braket{\epsilon_r | A_\lambda^{(l)} | \epsilon_m} = \iu\hslash \sum_{k = 1}^{l} \alpha_k {(\epsilon_r - \epsilon_m)}^{2k-1} \braket{\epsilon_r | \partial_\lambda \ham_0 | \epsilon_m},
		\end{equation}
		to be compared with the exact gauge potential:
		\begin{equation}
			\braket{\epsilon_r | A_\lambda | \epsilon_m} = \iu \hslash \frac{\braket{\epsilon_r | \partial_\lambda \ham_0 | \epsilon_m}}{\epsilon_m - \epsilon_r}.
		\end{equation}
		The proposed variational ansatz is a power-series approximation of the factor $ (\hslash\omega_{mr})^{-1} \equiv {(\epsilon_m - \epsilon_r)}^{-1} $ and will generally fail near $ \omega_{mr} = 0 $.
		
		Compared with the ``heuristic'' approach presented above, this NC ansatz has the obvious advantage that it can be improved by including higher-order terms and more variational parameters. The downside is that operators $ O_{2k-1} $ are nonlocal and as difficult to implement as the original proposal by~\textcite{Berry2009}. As opposed to the exact CD potential, however, these operators are well-defined around the quantum critical point.
		
		In this manuscript, we apply the counterdiabatic scheme to the ferromagnetic $ p $-spin model. 
        In the case $p=1$,	we only need $l=1$ (l is the order of nested commutators) to recover the exact CD potential.        
		For $ l = 1 $, the quadratic action $ S_1(\alpha) $ [see Eq.~\eqref{eq:action-quadratic}] is trivially minimized by
		\begin{equation}
		\alpha\ped{m} = -\frac{\Tr[O_0 O_2]}{\Tr\bigl[O_2^2\bigr]}.
		\end{equation}
		For the $ p $-spin system of Eq.~\eqref{eq:pspin-p-1}, $ O_2 = -8(1-\lambda) S_z + 8\lambda S_x $, thus, in the symmetric sector,
		\begin{align}
		\Tr[O_0 O_2] &= 16\Tr[S_z^2 + \lambda(S_x^2 - S_z^2) - S_x S_z ] \notag \\
		& = 16\sum_{i=0}^{\nspin} {\left(\frac{\nspin}{2} - i\right)}^2 = \frac{16\nspin (\nspin + 1) (\nspin + 2)}{12}.
		\end{align}
		In the same way, it is possible to prove that
		\begin{equation}
		\Tr\bigl[O_2^2\bigr] = \frac{64\nspin (\nspin + 1) (\nspin + 2)}{12} \bigl(1-2\lambda + 2 \lambda^2\bigr),
		\end{equation}
		so that
		\begin{equation}
		\alpha\ped{m} = -\frac{1}{4-8\lambda + 8 \lambda^2},
		\end{equation}
		independently of the system size.
		
		\section{Cyclic Ansatz}
		\label{CA_appendix}
		In Eq.~\eqref{eq:sy-ansatz}, we propose a general form for the CA. In this section, we will work out explicitly all the relevant terms in the case $ p = 3 $. 
		Starting from Eq.~\eqref{eq:sy-ansatz},
        we first explicitly evaluate all the terms:
        \begin{align}
        A_\lambda\api{CA} = \alpha_1 S_y +  \alpha_3 S_y^3 &+ \alpha_{xyz} S_x S_y S_z  - \alpha_{zyx} S_z S_y S_x \notag\\ 
         &+ \alpha_{yzx} S_y S_z S_x  - \alpha_{xzy} S_x S_z S_y  \notag\\ 
         &+ \alpha_{zxy} S_z S_x S_y  - \alpha_{yxz} S_y S_x S_z.
        \end{align} 
		Then, using the angular momentum commutation rules, we can simplify the previous expression to
		\begin{equation}
		A_\lambda\api{CA} = \alpha_1 S_y +  \alpha_3 S_y^3 + \alpha'  (S_x S_y S_z + \hc),
		\end{equation}  
		where we have defined $\alpha' = \alpha_{xyz}= - \alpha_{zyx} = - \alpha_{yxz}=  \alpha_{zxy}= - \alpha_{xzy}=  \alpha_{yzx}$ and we have used the fact that $(S_x S_y S_z+ \hc) = (S_z S_x S_y+\hc)=(S_x S_z S_y+\hc)$.
		With this assumption, the CA  requires just three variational parameters.

	\section{Whole Hilbert space versus maximum spin subspace: $ p = 3 $}
\label{app:whole-vs-maximum}
		
		\begin{figure*}[tb]
			\centering
			\includegraphics[width = 0.49\linewidth]{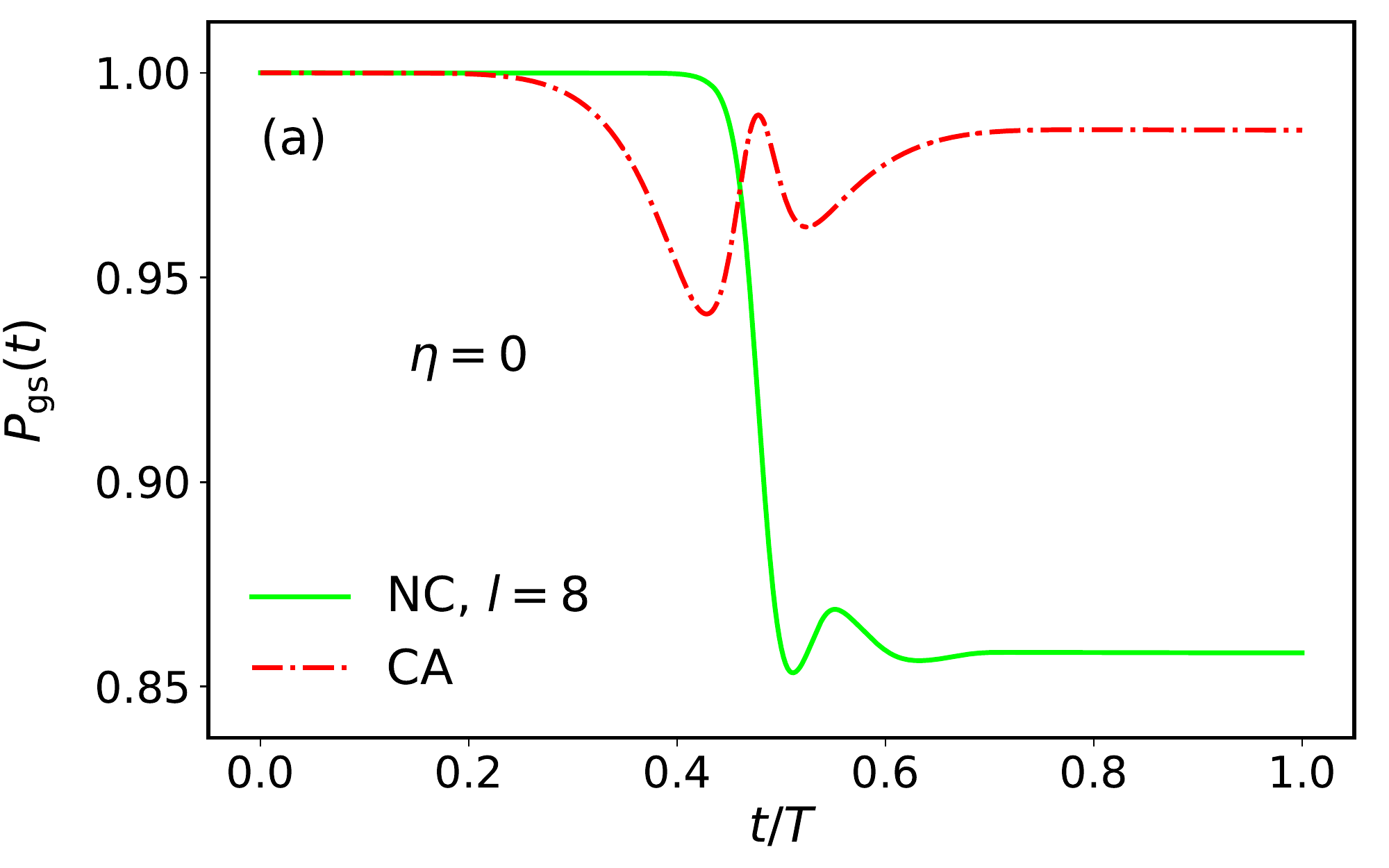}
			\includegraphics[width = 0.49\linewidth]{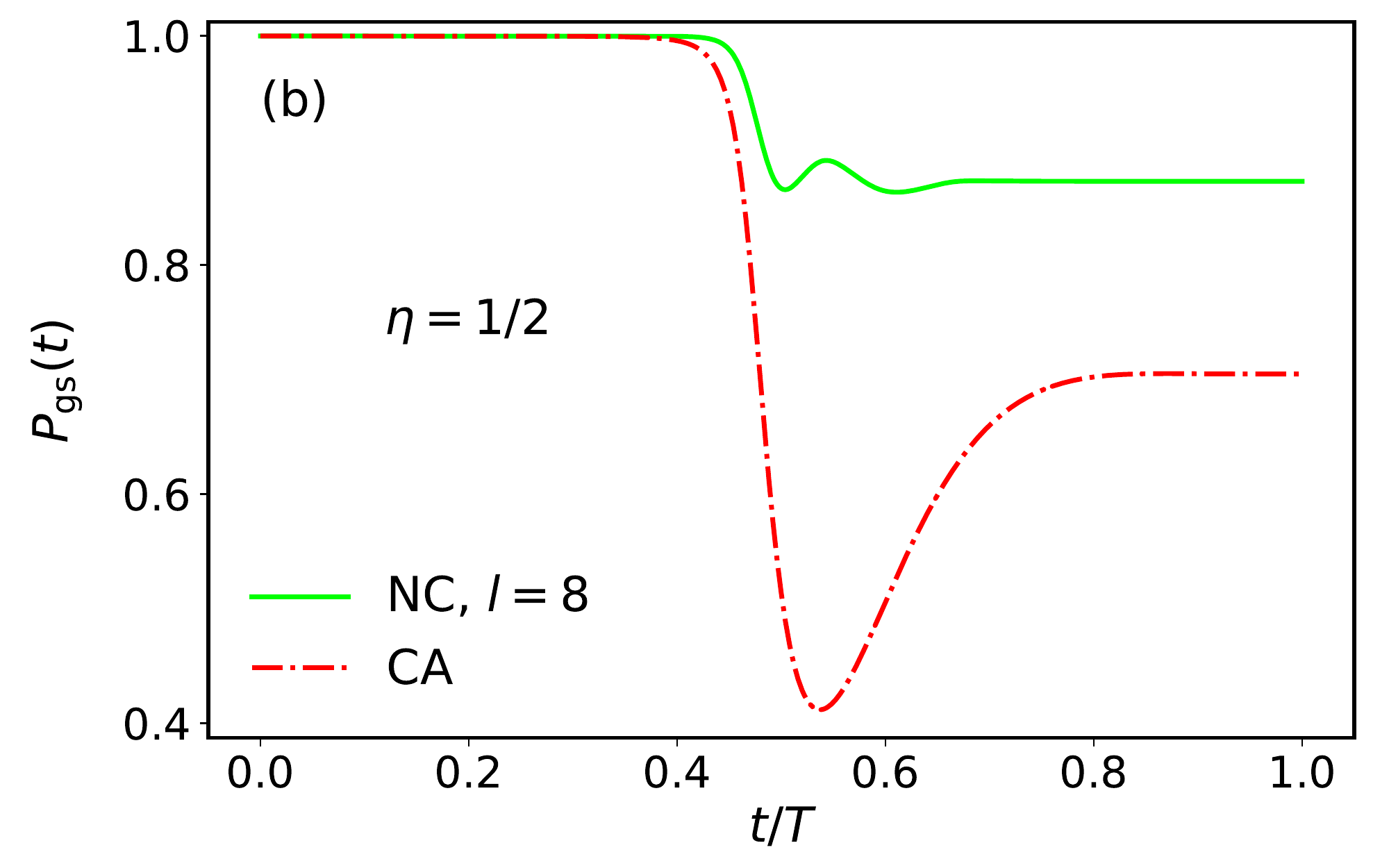}
			\caption{Ground state probability $ \pgs(t) $ as a function of time, for $ \nspin = 8 $ and $ p = 3 $. The annealing time is $ \tf = 1 $. Panel (a) shows the results obtained by the minimization of the action in Eq.~\eqref{eq:action-quadratic} in the symmetric subspace ($ \eta = 0 $). Panel (b) refers to the minimization in the whole Hilbert space ($ \eta = 1/2 $). In both panels, the green line refers to  nested commutator ansatz ($l = 8$), while the red dotdashed line refers to the CA.}
			\label{fig:n-8-p-3-symm-nosymm}
		\end{figure*}
		
		In this appendix, we compare the efficiency of our algorithms both for $\eta=0$ and $\eta=1/2$, \ie, calculating the traces in the maximum spin subspace or in the whole Hilbert space, respectively. In Fig.~\ref{fig:n-8-p-3-symm-nosymm}, we show the instantaneous ground state probability as a function of time, for $ \tf = 1 $ and $ \nspin = 8 $. The green line is for the NC ansatz of orders $ l = 8 $ and the red dotdashed line is the CA. The left (right) panel refers to the case $ \eta = 0 $ ($ \eta = 1/2 $).
		
Comparing the two panels, we notice that the ground state probability for the nested commutator ansatz is almost independent of $ \eta $. By contrast, the CA outcome is strongly affected by $ \eta $. In fact, for $ \eta = 0 $, the fidelity is very close to $ F = 1 $, while, for $ \eta = 1/2 $, the fidelity drops to $ F \approx \num{0.7} $. Also the instantaneous value of $ \pgs(t) $ has a different behavior depending on $ \eta $. In fact, for $ \eta = 0 $, the fidelity drops before the avoided crossing and then immediately recovers following a nonmonotonic behavior, with a maximum at the avoided crossing (see also Fig.~\ref{fig:p-3-dynamics}). By contrast, for $ \eta = 1/2 $, the ground state probability shows a minimum after the avoided crossing and then grows, with no maximum. Similar results also hold for other values of the system size $ \nspin $.
		

%

\end{document}